\documentstyle[12pt]{article}

\begin{document}
\begin{center}
{\large {\bf{PION AND KAON DISSOCIATION IN HOT QUARK MEDIUM}}}
\vskip 36pt
{\bf Abhijit Bhattacharyya$^a,$\footnote{Electronic Mail : 
abhijit@veccal.ernet.in}, Sanjay K. Ghosh$^b,$\footnote{Electronic Mail : 
phys@boseinst.ernet.in} and Sibaji Raha$^b,$\footnote{Electronic 
Mail : sibaji@boseinst.ernet.in}} 
\vskip 10pt
$^a$ Variable Energy Cyclotron Centre, 1/AF Bidhannagar,\\ 
Calcutta- 700 064, India\\
$^b$Department of Physics, Bose Institute, 
93/1, A.P.C.Road\\
Calcutta 700 009, India
\end{center}
\vskip 24pt
\begin{abstract}
Pion and kaon dissociation in a medium of hot quark matter is studied
in the Nambu Jona-Lasinio model.
The decay width of pion and kaon are found to be large but finite at 
temperatures much higher than the so called critical temperature of 
chiral or deconfinement transition, kaon decay width being larger.
Consequently, pions and even kaons (with a lower density compared to pions)
should coexist with quarks and gluons at such high temperatures.
The implication of the above result in the study of Quark-Gluon plasma is 
discussed. 
\end{abstract}
\newpage
\indent
A strong prediction of Quantum Chromodynamics (QCD), {\it the underlying 
theory of
strong interaction}, is that at very high temperature and/or density, the
bulk properties of strongly interacting matter would be governed by the
quarks and gluons, rather than the usual hadrons. Such a phase is called
quark gluon plasma (QGP) \cite{a} in the literature and the search for 
such a novel phase of matter constitutes a major area of current research 
in the field of high energy physics.
\par
The properties and dynamics of QGP are obviously governed by QCD. This 
conceptually 
straight forward task is, however, quite formidable in practice, particularly
because of the failure of perturbative QCD already in the temperature
range in the vicinity of $\Lambda_{QCD}$ ($\sim$ few hundred MeV) \cite{a1}. 
Analytical non-perturbative methods are not yet sufficiently developed
to be of much use in this context and as such, the lattice  formulation of
QCD has developed into the primary vehicle for the study of
QGP \cite{c}. In addition to the intensive 
computation, both in
terms of CPU time and numerical complexities, one can only address static
properties in the lattice. As a result, the space - time evolution of
the system formed in the ultrarelativistic heavy ion collisions remains
unapproachable in the framework of the lattice; thus the alternate, classical
picture of hydrodynamic evolution, which accounts for the overall energy -
momentum conservation in a collective manner and not much else, has been
used quite extensively to study the evolution of the QGP \cite{d}.
QCD inputs enter into such a picture through the equation of state 
of the QGP, preferably evaluated on the lattice ( but more often, through
a phenomenological bag model \cite{b}). 
\par
An inescapable feature of the collision
process is that the quarks and gluons must, at some epoch, turn into 
hadrons which would ultimately be detected, never the individual 
quarks and gluons. 
The actual process of hadronisation, however, continues to elude us. It has 
been 
widely postulated that there could be an actual phase transition (the order 
of which is an open issue), separating the QGP phase from the hadronic
phase \cite{f}. The recent results, showing the lack of thermodynamic 
equilibrium \cite{g} in the quark-gluon phase in ultrareletivistic heavy
ion collisions, indicate that such an ideal situation is unlikely. It
should also be noted at this juncture that although the persistence of
non-perturbative effects till very high temperatures was suggested in the 
literature quite early on \cite{h}, it is only recently that the lattice
results have confirmed that non-perturbative hadron like excitations 
could survive at temperatures far above the chiral phase transition 
temperature \cite{i}. The lattice result for pion screening mass has
been studied in ref. \cite{j}. The analysis of \cite{j} has been contradicted 
by Boyd {\it et al.} in ref. \cite{ja}. The conclusion of these 
authors \cite{ja}  
is consistent with the existence of free quarks 
at high temperatures. On the other hand, Shuryak \cite{jb} argued in a
subsequent work that the non-perturbative modes, especially pion- like
excitations, could indeed survive till temperatures above $T_{c}$.
Furthermore, similar results for pion screening 
masses are obtained in $\sigma$- model as well \cite{ja}.
It is thus imperative to understand the behaviour of
such hadronic resonances, their formation, stability and so on, in a
quark gluon medium at high temperature. In this work we confine
our attention to the case of pions and kaons only; these, being lighter  
than other hadrons, account for the bulk of the multiplicity. 
\par
Formation of light mesons like pions and kaons, a bound state of light 
relativistic quarks, is an
extremely difficult problem to handle in QCD. This is where all the
troublesome features of non-perturbative QCD would make their presence
felt. We therefore employ the usual 
practice of looking at the pion and kaons as a Goldstone boson arising from 
the spontaneous breaking of the chiral symmetry. In the present work
we have tried to understand the behaviour of pseudoscalar mesons  
($\pi$ and $K$) using Nambu Jona-Lasinio model \cite{k}. The decay width 
of pions and kaons have also been explored using the same model. 

The Nambu Jona-Lasinio (NJL) model, in its original form, was constructed 
as a pre-QCD theory of nucleons that interact via an effective two body 
interaction. This today is reinterpreted as a theory with quark degrees
of freedom. The Lagrangian density of this model incorporates the 
essential symmetries of QCD, the most important being the chiral symmetry.
The actual mechanism via which the chiral symmetry breaking occurs in the
NJL model follows closely the microscopic theory  of superconductivity.

The NJL model had its own shortcomings as well. The interaction between 
quarks is assumed to be point like and the model is non-renormalizable.
Furthermore, confinement is not incorporated in the NJL model; 
see ref. \cite{n}. Nevertheless, NJL model is very useful for our present
purpose, to understand the behaviour of pions and kaons (or more accurately, 
pion or kaon like excitations built out of quarks) in a hot quark medium.

The formulation of NJL model in flavour SU(3) was first introduced by
Hatsuda {\it et al.} \cite{l} and Bernard {\it et al.} \cite{m}.
The three flavour NJL model Lagrangian is written in terms of $u$, $d$ and
$s$ quarks, the interaction between them being constrained by the
$SU(3)_{L} \otimes SU(3)_{R}$ chiral symmetry, explicit symmetry breaking 
due to the current quark masses and the $U(1)_{A}$ breaking due to the axial 
anamoly \cite{n}. The full Lagrangian with KMT (Kobayashi- Maskawa
-'t-Hooft) term is given below \cite{n}.
\begin{eqnarray}
\cal {L} &=& \bar{q}(i\gamma\cdot\partial - {\bf m})q 
+ {1\over 2}g_{s} {\sum_{a=0}}^{8} [(\bar{q}\lambda_{a}q)^{2}
(\bar{q}i\lambda_{a}\gamma_{5}q)^{2}] \nonumber \\
&+& g_{D}[det \bar{q_i}(1-\gamma_{5})q_{j} + h.c.]
\label{eq:lnjl}
\end{eqnarray}
where the quark fields $q_{i}$ has three colours ($N_{c}=3$) and three
flavours ($N_{f}=3$) and $\lambda_{a}$ ($a= 1,8$) are the Gell-Mann 
matrices. The quark mass matrix is given by 
${\bf m}= diag(m_{u},m_{d},m_{s})$.

In the mean field approximation, the quark condensates at finite temperature
are given by \cite{n},
\begin{eqnarray}
<<\bar{q_{i}} q_{i}>>= - 2N_{c}\int {d^3p \over{2 \pi^3}} 
{M_{i} \over {E_{ip}}} f(E_{ip})
\label{eq:conden}
\end{eqnarray}
where $E_{ip}$ is the quark single particle energy for the i-th specie and 
$f(E_{ip})=1 - n_{ip} - \bar{n}_{ip}$, $n_{ip}$ and $\bar {n}_{ip}$ being 
the Fermi-Dirac distributions for quarks and anti-quarks. 
If quark chemical potential is zero, then 
$n_{ip}= \bar {n}_{ip} = [exp(E_{ip}/T)~~+~~1]^{-1}$.

The temperature dependent constituent quark masses $M_{i}$ are obtained
from the expressions below,
\begin{eqnarray}
M_{u}=m_{u} - 2g_{s}\alpha - 2g_{D}\beta\gamma   \nonumber \\
M_{d}=m_{d} - 2g_{s}\beta - 2g_{D}\alpha\gamma   \nonumber \\ 
M_{s}=m_{s} - 2g_{s}\gamma - 2g_{D}\alpha\beta    
\label{eq:conmas}
\end{eqnarray}
where
\begin{eqnarray}
<<\bar{u} u>> \equiv \alpha,~~~~<<\bar{d} d>> \equiv \beta, 
~~~~<<\bar{s} s>> \equiv \gamma
\label{eq:defalpha}
\end{eqnarray}

The actual formation of pions from quarks and
gluons would require an involved analysis through the Bethe-Salpeter
equation. Such a study is very much on our agenda but we do not address
this issue here. In the present work, we concern ourselves with the decay
of pionic and kaonic excitations, the properties of which we assume to be 
given by the NJL model. It should be reiterated that at
temperatures above the critical temperature, these mesonic excitations are
more like resonances with large effective masses \cite{i,j}. 
In the following, we study the decay width of such
pionic excitations in the hot quark medium as a function of temperature,
starting with the Lagrangian given above in equation (\ref{eq:lnjl}). 
\par
The quark mass $M_{i}$ appearing in eq. (\ref{eq:conden}) and 
in eq. (\ref{eq:conmas}) is a very important
ingredient in our calculation. In the absence of any medium and/or
dynamic effect, $M_{i}$ should assume the value of the current quark mass. 
On the other hand, we know that due to the spontaneous breakdown of the
chiral symmetry, quarks attain the value of the constituent quark mass 
\cite{k}.
\par
In the present calculation we have used the parametrisation of 
ref. \cite{m} ( $\Lambda$ = 631.4
, $g_{s}\Lambda^2$ = 3.67, $g_{D}\Lambda^5$ = -9.29 and 
current mass $m_{u,d}(m_s)$ = 5.5 (135.7) MeV ) to calculate the quark and 
meson masses. The constituent quark masses are calculated using the 
gap equations(eq. \ref{eq:conmas}). These quark masses are then put into 
the dispersion equation \cite{n} for mesons to get dynamical masses of 
mesons ($\pi$ and $K$, here).
\begin{eqnarray}
1~~ +~~ 2G_{\phi}\Pi_{ij}(\omega,\vec{q}\rightarrow 0)~~ =~~ 0
\end{eqnarray}
where $\Pi_{ij}$ is the one loop polarization due to $u$ and $d$ quark for 
pions and $u$ or $d$ and $s$ quark for kaons. $G_{\phi}$ is the coupling 
constant with $\phi$ coerresponding to $\pi$ or $K$. 
The general expression for polarization function is 
\begin{eqnarray}
\Pi(q_{0},\vec{q})={N_{c} \over {(2\pi)^3}} {\int_{0}}^{\Lambda} 
{d^3 p \over E_{p} E_{k}}
\left[ (n_{k} - n_{p}) \left\{ {1 \over {E_{p}-E_{k}+q_{0}+i\epsilon}}
+ {1 \over {E_{p}-E_{k}-q_{0}-i\epsilon}} \right\} \right.  \nonumber \\
\times (-E_{p}E_{k} + \vec{p}.\vec{k} + M_{1} M_{2}) \nonumber \\
 + (n_{k} + n_{p}-1) \left\{ {1 \over {E_{p}+E_{k}+q_{0}+i\epsilon}}
+ {1\over {E_{p}+E_{k}-q_{0}-i\epsilon}}\right\}   \nonumber \\
\left. \times (E_{p}E_{k} + \vec{p}.\vec{k} + M_{1} M_{2}) \right]
\label{eq:polar}
\end{eqnarray}
where $N_{c}$ is the number of colours and $\vec{k} = \vec{p} + \vec{q}$.
The energies $E_{p}=\sqrt{p^2+{M_{1}}^2}$ and 
$E_{k}=\sqrt{(\vec{p}+ \vec {q})^2+ {M_{2}}^2}$. 
For pion, $M_{1} = M_{2} = M_{u}$. For kaon, 
$M_{1}= M_{u(d)}$ and $M_{2} = M_{s}$. $n_{k}$ and $n_{p}$ are the 
Fermi-Dirac distribution functions defined earlier.
The pseudoscalar couplings are,
\begin{eqnarray}
G_{\pi}=g_{s}~~+~~g_{D} \gamma \nonumber \\
G_{K^{\pm}}=g_{s}~~+~~g_{D} \beta \nonumber \\
G_{K^{0}}=g_{s}~~+~~g_{D} \alpha
\end{eqnarray}
where $\alpha$, $\beta$ and $\gamma$ are defined in eq. (\ref{eq:defalpha}).

The decay width is evaluated using the imaginary part of the 
eq.(\ref{eq:polar}) as given below,
\begin{eqnarray}
\Gamma_{\phi}=-{{G_{\phi q}}^2 Im\Pi(\omega,\vec{q}\rightarrow 0) \over 
\omega}
\label{eq:decnjl}
\end{eqnarray}
where $G_{\phi q}$ is the empirical meson-quark coupling as obtained in NJL.
Here we have used $G_{\pi q} = 3.5$ and $G_{K q} = 3.6$\cite{m}.

The variation of quark and meson masses is shown in figure 1. The
$u$ or $d$ quark masses starting from 135 MeV drops to the current quark mass
value just after a temperature of 200 MeV. On the other hand, the drop in
the strange quark mass is much smaller around that temperature, showing the 
effect of explicitly broken chiral symmetry, by a larger amount, in the SU(3) 
sector. Pion and kaon both show
a similar qualitative behaviour. The masses of pion and kaon remain constant 
at their free masses upto a temperature 200 MeV but increases sharply after 
that with a pion mass of 900 MeV and Kaon mass of 100 MeV around 450 MeV 
temperature, thus giving a slower increment for Kaons compared to pions.
The difference in the behaviour of Kaon and pion  can be attributed to
the difference in the behaviour of $u$ and $s$ quark masses. 

The difference in the behaviour of pion and kaon is more prominent in the 
decay width, as shown in figure 2.
Figure 2 shows that the decay width is very high at high temperature 
and decreases with decreasing temperature, going to zero 
at around $T = 0.2$ GeV. It is worth noticing that at around the same 
temperature, the effective pion mass attains the value of the free pion 
mass (figure 1). The decay width of Kaon is around 3 GeV where as that of 
pion is around 1.4 GeV at 500 MeV temperature. This is a very significant 
result due to two reasons. Firstly, our results show that though there
will be pions and kaons along with the quarks at high temperature phase, 
the numbers of mesons will be very small due to their large decay width.
Moreover, the number of kaons will be much less compared to pions at high
temperature phase, though both the mesons will become stable around the same
temperature (below 200 MeV temperature).

\par
Our results will have a strong bearing on the study of hadronisation. As 
already mentioned, the lack of thermodynamic equilibrium in a QGP system 
implies that one may not get a clear-cut phase transition from QGP to 
hadrons. Thus, to understand the process of hadronisation, one should 
really start from a very high temperature ($>>$ expected $T_c$) and 
then let the system evolve dynamically towards  lower temperatures. 
Here what one would find, as indicated from our present calculation, 
is that initially a very small number of 
pions and kaons would be present in the system along with quarks and gluons.  
Then, even if additional mesons are formed through $q {\bar q}$ fusion 
and/or bound state formation, the total number of mesons would not increase 
very fast, as most of them must decay immediately due to the large 
decay width at such high temperatures. Only in the 
vicinity of $T \sim 200 MeV$, where the decay width is small, the number 
of mesons would start increasing significantly and gradually become dominant 
compared to the number of quarks at some lower temperature. However, the  
exact value of the temperature, at which the decay width goes to zero, 
 will depend on the value of the quark mass considered.
\par
To summarise, we have calculated, for the first time, the decay of pions and
kaons, in a hot quark medium. The most
interesting and noteworthy feature is that, even without any consideration
of the detailed evolution and dynamics of the system, the pionic modes are
found to dominate around a temperature of $200$ MeV. Though the question 
whether 
this is a signature of a phase transition cannot be addressed within the
framework of the present work, the fact that most of the pions and kaons
decay into quarks, owing to a large decay width at temperatures higher than 
$T=200$ MeV, is a remarkable finding. Moreover, both the pionic as well 
as kaonic modes start becoming important at about the same temperature, thus
providing a hint of some kind of a transition temperature.
\par
The work of AB and SKG have been supported, in part, by the Department of 
Atomic Energy (Government of India) and Council of Scientific and Industrial 
Research (Government of India), respectively.  


\begin{thebibliography}{99}
\bibitem{f} It should be noted that, in the context of QGP or hot hadronic
matter, one comes across two kinds of phase transitions, namely, the chiral 
phase transition and the deconfinement phase transition. The current consensus 
is that although these two phase transitions are certainly distinct from 
each other, the critical temperatures may not be much different. We have taken 
them to be same for the present purpose.
\end{thebibliography}
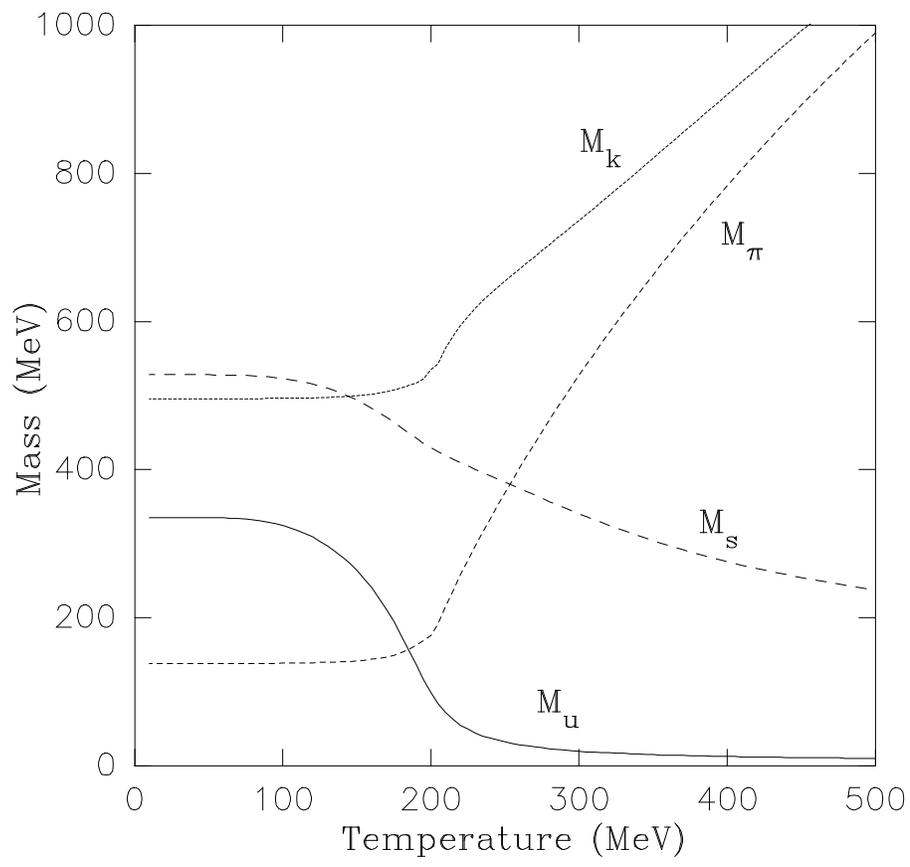
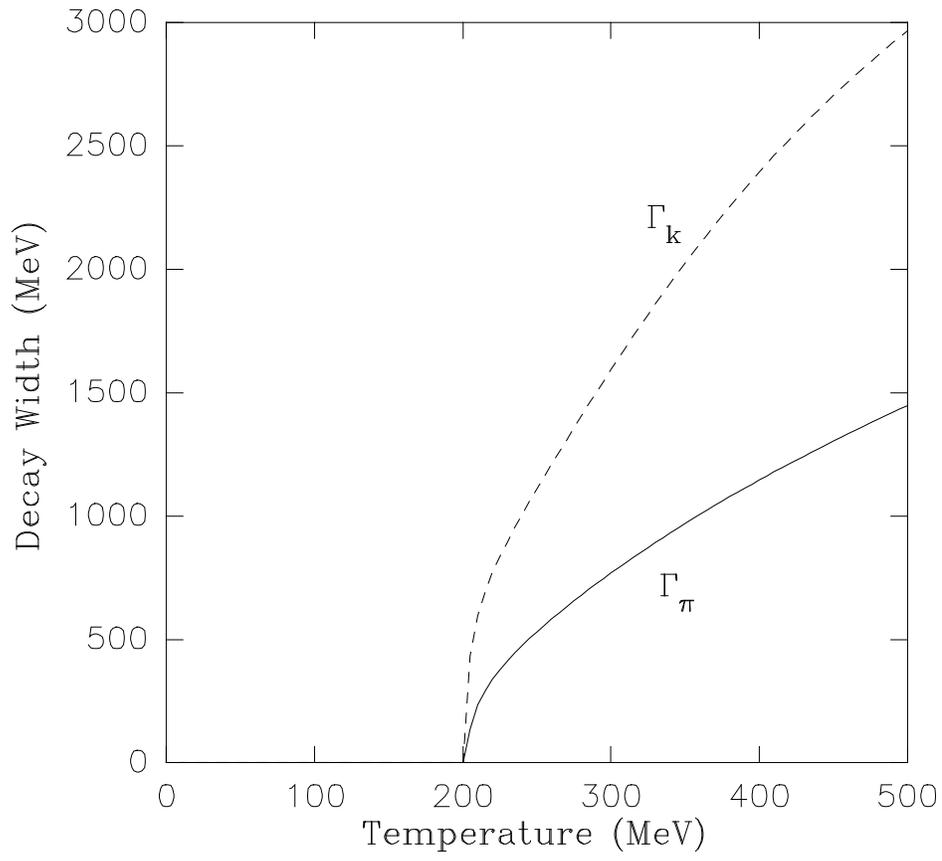
\end{document}